\documentclass[a4paper]{jpconf}
\usepackage{graphicx}

\def\lsim{\mathrel{\raise.3ex\hbox{$<$\kern-.75em\lower1ex\hbox{$\sim$}}}}
\def\gtrsim{\mathrel{\raise.3ex\hbox{$>$\kern-.75em\lower1ex\hbox{$\sim$}}}}

\begin{document}

\title{The Indirect Search for Dark Matter with IceCube}

\author{Francis Halzen$^a$, Dan Hooper$^{b,c}$} 
\address{$^a$ University of Wisconsin, Madison, USA\\
$^b$ Fermi National Accelerator Laboratory, USA\\
$^c$ University of Chicago, USA}

\section{The First Kilometer-Scale High Energy Neutrino Detector: IceCube}
\vspace{.2cm}
A series of first-generation experiments~\cite{Spiering:2008ux} have demonstrated that high energy neutrinos with $\sim10$\,GeV energy and above can be detected by observing the Cherenkov radiation from secondary particles produced in neutrino interactions inside large volumes of highly transparent ice or water instrumented with a lattice of photomultiplier tubes. The first second-generation detector, IceCube, is under construction at the geographic South Pole~\cite{Halzen:2006mq}. IceCube will consist of 80 kilometer-length strings, each instrumented with 60 10-inch photomultipliers spaced by 17 m. The deepest module is located at a depth of 2.450\,km so that the instrument is shielded from the large background of cosmic rays at the surface by approximately 1.5\,km of ice. The strings are arranged at the apexes of equilateral triangles 125m on a side. The instrumented detector volume is a cubic kilometer of dark, highly transparent and totally sterile Antarctic ice. The radioactive background is dominated by the instrumentation deployed into the natural ice. A surface air shower detector, IceTop, consisting of 160 Auger-style 2.7m diameter ice-filled Cherenkov detectors deployed pairwise at the top of each in-ice string, augments the deep-ice component by providing a tool for calibration, background rejection and cosmic ray studies.

Each optical sensor consists of a glass sphere containing the photomultiplier and the electronics board that digitizes the signals locally using an on-board computer. The digitized signals are given a global time stamp with residuals accurate to less than 3\,ns and are subsequently transmitted to the surface. Processors at the surface continuously collect the time-stamped signals from the optical modules that each function independently.  The digital messages are sent to a string processor and a global event trigger. They are subsequently sorted into the Cherenkov patterns emitted by secondary muon tracks that reveal the direction of the parent neutrino. In this context we will focus on the neutrino flux from the direction of the Sun and the Earth.

IceCube detects neutrinos with energies in excess of 0.1\,TeV. An upgrade of the detector, dubbed Deep Core, consists of an infill of 6 strings with 60 DOMs with high quantum efficiency. They are mostly deployed in the highly transparent ice making up the bottom half of the IceCube detector. Deep Core will decrease the threshold to $\sim10$\,GeV over a significant fraction of IceCube's fiducial volume and will be complete by February 2010; see Fig\,1.
\begin{figure}
\begin{center}
\includegraphics[width=5.0in,angle=0]{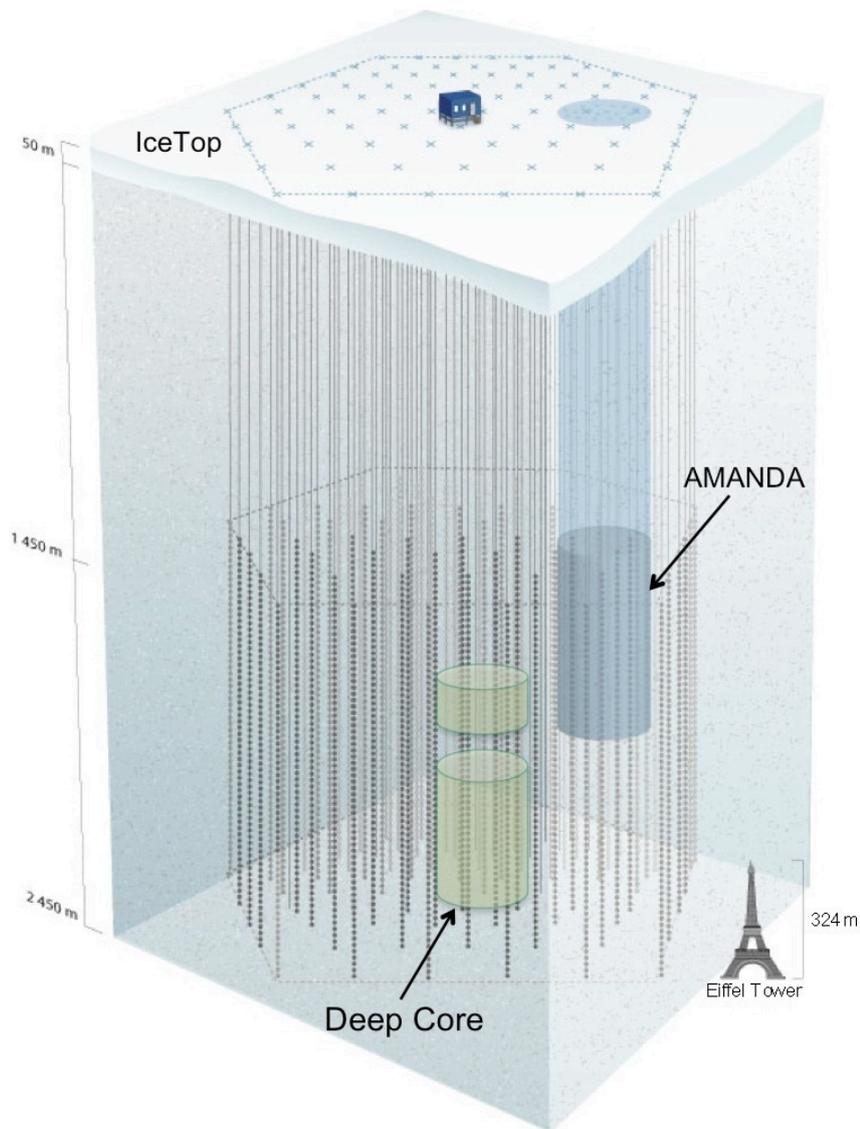}
\end{center}
\caption{The IceCube detector, consisting of IceCube and IceTop and the low-energy sub-detector DeepCore.}
\end{figure}

The main scientific goals of IceCube fall into broad categories:
\begin{enumerate}
\item Detect astrophysical neutrinos produced in cosmic sources with an energy density comparable to the energy density in cosmic rays~\cite{Gaisser:1994yf}. Supernova remnants satisfy this requirement if they are indeed the sources of the galactic cosmic rays as first proposed by Zwicky; his proposal is a matter of debate after more than seventy years. The sources of the extragalactic cosmic rays naturally satisfy the prerequisite when particles accelerated near black holes, possibly the central engines of active galaxies or gamma ray bursts, collide with photons in the associated radiation fields. While the secondary protons may remain trapped in the acceleration region, approximately equal numbers of neutrons, neutral and charged pions escape. The energy escaping the source is therefore distributed between cosmic rays, gamma rays and neutrinos produced by the decay of neutrons and neutral and charged pions, respectively.
\item As for conventional astronomy, neutrino astronomers observe the neutrino sky through the atmosphere. This is a curse and a blessing; the background of neutrinos produced by cosmic rays in interactions with atmospheric nuclei provides a beam essential for calibrating the instrument. It also presents us with an opportunity to do particle physics~\cite{GonzalezGarcia:2005xw}. Especially unique is the energy range of the ÒbackgroundÓ atmospheric neutrinos covering the range $0.1 - 10^5$\,TeV, energies not within reach of accelerators. Cosmic beams of even higher energy may exist, but the atmospheric beam is guaranteed. IceCube is expected to collect a data set of order one million neutrinos over ten years with a scientific potential that is only limited by our imagination. With the Deep Core upgrade, IceCube will accumulate atmospheric neutrinos with sufficient statistics to determine the flavor hierarchy by observing the matter effect of the atmospheric neutrino beam near the first oscillation dip just below 10\,GeV. A positive result will require a sufficient understanding of the systematics and, of course, a mixing angle $\theta_{13}$ that is not too small. Neither one is guaranteed at this point but the good news is that the relevant data are forthcoming in the next few years.
\item The passage of a large flux of MeV-energy neutrinos produced by a galactic supernova over a period of seconds will be detected as an excess of the background counting rate in all individual optical modules~\cite{Halzen:1995ex}. Although only a counting experiment, IceCube will measure the time profile of a neutrino burst near the center of the Galaxy with the statistics of about one million events, equivalent to the sensitivity of a 2 megaton detector.
\item Finally, the subject of this chapter, IceCube will search for neutrinos from the annihilation of dark matter particles gravitationally trapped at the center of the Sun and the Earth\cite{original}.
\end{enumerate}

These scientific missions have historically set the scale of the detector. In this context it is worthy of note that the AMANDA detector, the forerunner and proof of concept for IceCube, received a significant fraction of its initial funding from the Berkeley Center for Particle Astrophysics to search for dark matter.

\section{Indirect Search for Dark Matter with Neutrinos: Status}
\vspace{.2cm}
The evidence that yet to be detected weakly interacting massive particles (WIMPs) make up the dark matter is compelling. WIMPs are swept up by the Sun as the Solar System moves about the galactic halo. Though interacting weakly, they will ocassionally scatter elastically with nuclei in the Sun and lose enough momentum to become gravitationally bound. Over the lifetime of the Sun, a sufficient density of WIMPs may accumulate in its center so that an equilibrium is established between their capture and annihilation. The annihilation products of these WIMPs represent an indirect signature of halo dark matter, their presence revealed by neutrinos which escape the Sun with minimal absorption. The neutrinos are, for instance, the decay products of heavy quarks and weak bosons resulting from the annihilation of  WIMPs into $\chi\chi\rightarrow b\bar{b}$ or $W^+ W^-$. These can be efficiently identified by the neutrino detectors discussed above because of the relatively large neutrino energy of order of the mass of the WIMP.

The beauty of the indirect detection technique using neutrinos is that the astrophysics of the problem is understood. The source in the sun has built up over solar time sampling the dark matter throughout the galaxy; therefore, any possible structure in the halo has been averaged out. Given a WIMP mass and properties, one can unambiguously predict the signal in a neutrino telescope. If not observed, the model is ruled out. This is in contrast to indirect searches for photons from WIMP annihilation, whose sensitivity depends critically on the structure of halo dark matter; observation requires cuspy structure near the galactic center or clustering on appropriate scales elsewhere. Observation not only necessitates appropriate WIMP properties, but also favorable astrophysical circumstances. 

Extensions of the Standard Model of quarks and leptons, required to solve the hierarchy problem, naturally yield dark matter candidates. For instance, the neutralino, the lightest stable particle in supersymmetric models, has been intensively studied as a possible dark matter candidate. Detecting it has become the benchmark by which experiments are evaluated and mutually compared. We will follow this tradition here. Supersymmetric models allow for a large number of free parameters and, unfortunately, for a variety of parameter sets that are able to generate the observed dark matter density in the context of standard big bang cosmology. How to meaningfully sample this parameter space is not straightforward. In this chapter we will first demonstrate the considerable potential of neutrino telescopes for detecting dark matter based on a set of very general models. These models satisfy all experimental and cosmological constraints but do not assume a particular form of symmetry breaking. Subsequently, we will revisit IceCube's sensitivity based on a more constrained set selected for assessing the performance of the LHC; the experimental input has been updated~\cite{Allanach:2008iq}.

The neutralino interacts with ordinary matter by spin-independent (e.g. Higgs exchange) and by spin-dependent (e.g. Z-boson exchange) interactions. The first mechanism favors direct detection experiments~\cite{Sadoulet:2007pk} because the WIMP interacts coherently, resulting in an increase in sensitivity proportional to the square of the atomic number of the detector material. Although competitive now for larger WIMP masses, indirect detection experiments are unlikely to remain competitive in the future given the rapid improvement of the sensitivity of direct experiments. We will show that this is not the case for spin-dependent models, however. Within the context of supersymmetry, direct and indirect experiments are complementary.

Although IceCube detects neutrinos of all flavors, sensitivity to neutrinos produced by WIMPs in the sun is achieved by exploiting the degree accuracy with which muon neutrinos can be pointed back to the Sun. Data taken with the first 22 strings of IceCube has resulted in a limit on an excess flux from the Sun shown in Fig.\,2. This figure shows the current limits on the neutrino induced muon flux from the direction of the Sun, as well as the projected sensitivity of IceCube. Also shown is a sampling of the supersymmetric parameter space. The current limit improves on previous results of the Super-Kamiokande~\cite{Desai:2004pq} and AMANDA~\cite{Ackerman:2005fr} collaborations by factors of 3 to 5 for WIMPs heavier than approximately 250 GeV. The current IceCube limit is transformed into a limit on the spin-dependent cross section in Fig\,3. Forthcoming analyses using the complete AMANDA data set will improve on the result shown and will probe the model space shown in Fig.\,2 for masses below 250\,GeV. Deep Core is under construction and will enable IceCube to place the strongest limits to-date for WIMPs as light as 50 GeV. Results from the Antares collaboration obtained with data from the recently completed detector, are forthcoming\cite{antares}. For WIMPs lighter than approximately 50 GeV, the reach of Super-Kamiokande is unmatched by existing detectors. 

\begin{figure}
\begin{center}
\includegraphics[width=4in,angle=0]{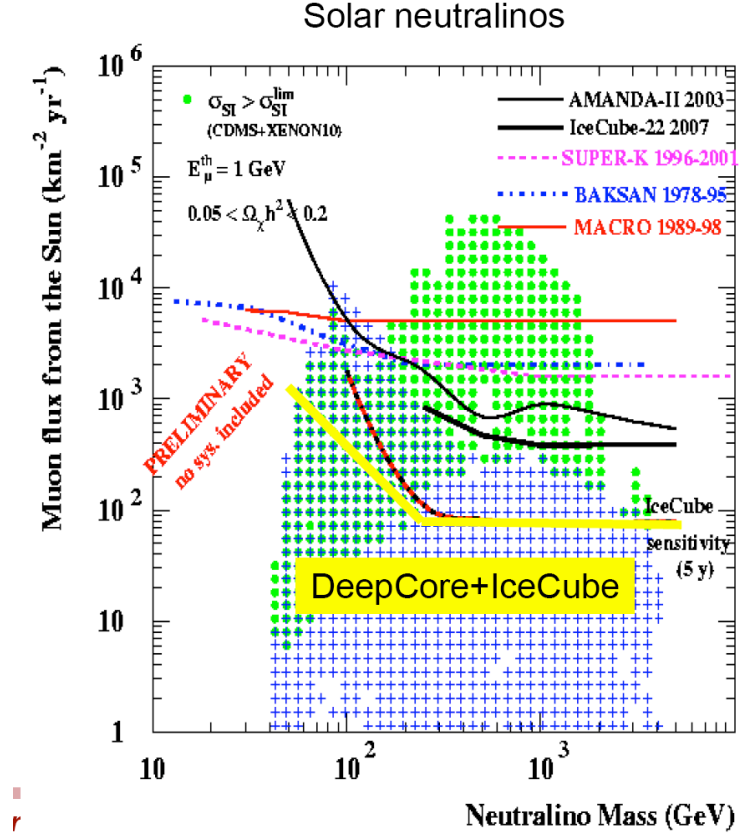}
\end{center}
\caption{The current status of indirect searches for dark matter using high energy neutrinos. In addition to the various experimental constraints, the projected sensitivity of IceCube is shown, along with the flux of neutrino induced muons predicted over the parameter space of the Minimal Supersymmetric Standard Model~\cite{IC22}. Muons, simulated with energies in excess of 1~GeV, are required to trigger the detector. }
\end{figure}

\begin{figure}
 \begin{center}
\includegraphics[width=4.0in,angle=270]{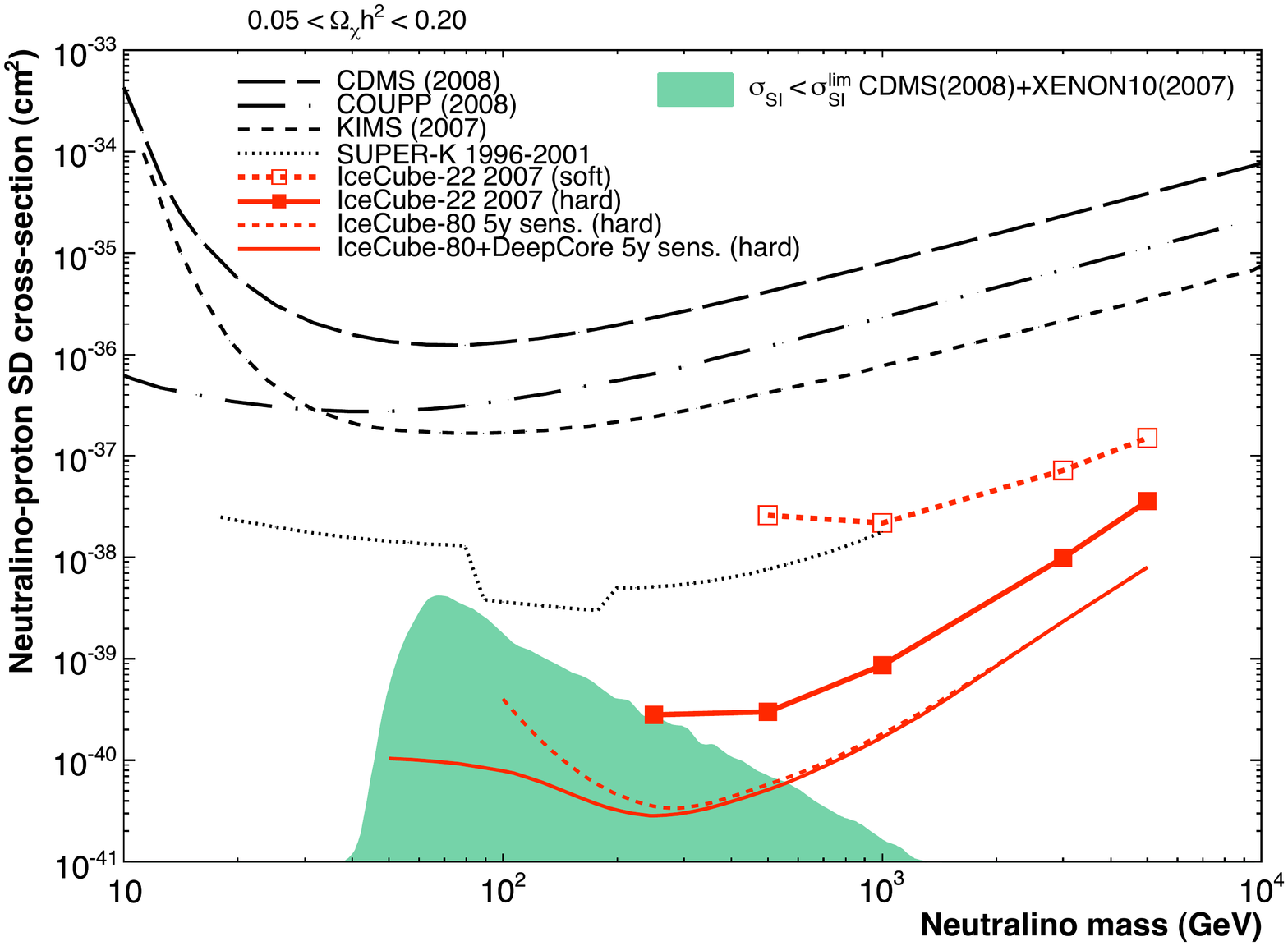}
\end{center}
\caption{Upper limits at 90\% confidence level on the spin-dependent neutralino-proton cross section assuming that the neutrinos are produced by $b\bar{b}$ and WW annihilation~\cite{IC22}. The limits have been obtained with IceCube operated with 22 out of 80 strings. Also shown is the reach of the completed detector. Limits from the Super-Kamiokande and direct detection experiments are shown for comparison. The shaded region represent supersymmetric models not ruled out by direct experiments.}
\end{figure}

In the remainder of this chapter, we describe the processes of capture and annihilation of WIMPs in the Sun, and predict the resulting flux of high energy neutrinos. In the later sections of this chapter, we will discuss the sensitivity of this technique to neutralino dark matter. We will also comment at the end on the excellent sensitivity of IceCube to Kaluza-Klein dark matter candidates.

\section{The Capture and Annihilation of WIMPs in the Sun}
\vspace{.2cm}
Beginning with a simple estimate, we expect WIMPs to be captured in the Sun at a rate approximately given by:

\begin{equation}
C^{\odot} \sim \phi_{\chi} (M_{\odot}/m_p) \sigma_{\chi p},
\end{equation}
where $\phi_{\chi}$ is the flux of WIMPs in the Solar System, $M_{\odot}$ is the mass of the Sun, $\sigma_{\chi p}$ is the WIMP-proton elastic scattering cross section and $m_p$ the proton mass. Reasonable estimates of the local distribution of WIMPs leads to a capture rate of $C^{\odot} \sim 10^{20} \, {\rm sec}^{-1} \times (100 \, {\rm GeV}/m_{\chi})\, (\sigma_{\chi p}/10^{-6}\, {\rm pb})$, where $m_{\chi}$ is the mass of the WIMP. This neglects a number of potentially important factors, including the gravitational focusing of the WIMP flux toward the Sun, and the fact that not every scattered WIMP will ultimately be captured. Taking these effects into account leads us to a solar capture rate of~\cite{Gould:1991hx}: 
\begin{eqnarray}
C^{\odot} \approx 1.3 \times 10^{21} \, \mathrm{sec}^{-1}  
\left( \frac{\rho_{\mathrm{local}}}{0.3\, \mathrm{GeV}/\mathrm{cm}^3} \right) 
\left( \frac{270\, \mathrm{km/s}}{\bar{v}_{\mathrm{local}}} \right)  \nonumber \\
\times \left( \frac{100 \, \mathrm{GeV}}{m_{\chi}} \right) \sum_i \left( \frac{A_i \, (\sigma_{\mathrm{\chi i, SD}} +\, \sigma_{\mathrm{\chi i, SI}}) \,S(m_{\chi}/m_{i})} {10^{-6}\, \mathrm{pb}} \right) ,
\label{capture}
\end{eqnarray}
where $\rho_{\mathrm{local}}$ is the local dark matter density and $\bar{v}_{\mathrm{local}}$ 
is the local rms velocity of halo dark matter particles. $\sigma_{\mathrm{\chi i, SD}}$ and $\sigma_{\mathrm{\chi i, SI}}$ are the spin-dependent and spin-independent elastic scattering cross sections of the WIMP with nuclei species $i$, and $A_i$ is a factor denoting the relative abundance and form factor for each species. In the case of the Sun, $A_{\rm H} \approx 1.0$, $A_{\rm He} \approx 0.07$, and $A_{\rm O} \approx 0.0005$. The quantity $S$ contains dynamical information and is given by:

\begin{equation}
S(x)=\bigg[\frac{A(x)^{3/2}}{1+A(x)^{3/2}}\bigg]^{2/3}
\end{equation}
where
\begin{equation}
A(x)=\frac{3}{2}\frac{x}{(x-1)^2}\bigg(\frac{v_{\rm esc}}{\bar{v}_{\mathrm{local}}}\bigg)^2,
\end{equation}

and $v_{\rm esc} \approx 1156 \, {\rm km/s}$ is the escape velocity of the Sun. Notice that for WIMPs much heavier than the target nuclei $S \propto 1/m_{\chi}$, leading the capture rate to be suppressed by two factors of the WIMP mass. In this case ($m_{\chi} \gtrsim 30$\,GeV), the capture rate can be approximated as:

\begin{eqnarray}
C^{\odot} \approx 3.35 \times 10^{20} \, \mathrm{sec}^{-1} 
\left( \frac{\rho_{\mathrm{local}}}{0.3\, \mathrm{GeV}/\mathrm{cm}^3} \right) 
\left( \frac{270\, \mathrm{km/s}}{\bar{v}_{\mathrm{local}}} \right)^3 \, \left( \frac{100 \, \mathrm{GeV}}{m_{\chi}} \right)^2 \nonumber \\ 
\times \left( \frac{\sigma_{\chi \mathrm{H, SD}} +\, \sigma_{\chi \mathrm{H, SI}}
+ 0.07 \, \sigma_{\chi \mathrm{He, SI}}+ 0.0005 \,S(m_{\chi}/m_{\mathrm{O}})\, \sigma_{\chi \mathrm{O, SI}}     } {10^{-6}\, \mathrm{pb}} \right).
\label{SIeff}
\end{eqnarray}

If the capture rate and annihilation cross sections are sufficiently large, equilibrium will be reached between these processes. The time dependence of the number of WIMPs $N(t)$ in the Sun is given by

\begin{equation}
\dot{N}(t) = C^{\odot} - A^{\odot} N(t)^2 - E^{\odot} N,
\end{equation}
where $C^{\odot}$ is the capture rate described above and $A^{\odot}$ is the 
annihilation cross section times the relative WIMP velocity per volume. $E^{\odot}$ is the inverse time for a WIMP to escape the Sun by evaporation. Evaporation is highly suppressed for WIMPs heavier than a few GeV~\cite{Gould:1987ir,Griest:1986yu}. $A^{\odot}$ can be approximated by

\begin{equation}
A^{\odot} = \frac{\langle \sigma v \rangle}{V_{\mathrm{eff}}}, 
\end{equation}
where $V_{\mathrm{eff}}$ is the effective volume of the core
of the Sun determined by matching the core temperature with 
the gravitational potential energy of a single WIMP at the core
radius. It is given by~\cite{Gould:1987ir,Griest:1986yu}

\begin{equation}
V_{\rm eff} = 5.7 \times 10^{27} \, \mathrm{cm}^3 
\left( \frac{100 \, \mathrm{GeV}}{m_{\chi}} \right)^{3/2} \;.
\end{equation}

Neglecting evaporation, the present WIMP annihilation rate is given by

\begin{equation} 
\Gamma = \frac{1}{2} A^{\odot} N(t_{\odot})^2 = \frac{1}{2} \, C^{\odot} \, 
\tanh^2 \left( \sqrt{C^{\odot} A^{\odot}} \, t_{\odot} \right) \;, 
\end{equation}
where $t_{\odot} \approx 4.5$ billion years is the age of the solar system.
The annihilation rate is maximized when it reaches equilibrium with
the capture rate.  This occurs when 

\begin{equation}
\sqrt{C^{\odot} A^{\odot}} t_{\odot} \gg 1 \; .
\end{equation}
If this condition is met, the final annihilation rate (and corresponding neutrino flux and event rate) is determined entirely by the capture rate and has no further dependence on the dark matter particle's annihilation cross section.

The muon neutrino spectrum at the Earth resulting from WIMP annihilations in the Sun is given by:

\begin{equation}
\frac{dN_{\nu_{\mu}}}{dE_{\nu_{\mu}}} = \frac{ C_{\odot} F_{\rm{Eq}}}{4 \pi D_{\rm{ES}}^2}   \bigg(\frac{dN_{\nu_{\mu}}}{dE_{\nu_{\mu}}}\bigg)^{\rm{Inj}},
\label{wimpflux}
\end{equation}
where $C_{\odot}$ is the WIMP capture rate in the Sun, $F_{\rm{Eq}}$ is the non-equilibrium suppression factor ($\approx 1$ for capture-annihilation equilibrium), $D_{\rm{ES}}$ is the Earth-Sun distance and $(\frac{dN_{\nu_{\mu}}}{dE_{\nu_{\mu}}})^{\rm{Inj}}$ is the muon neutrino spectrum from the Sun per WIMP annihilating. This result is modified by propagation effects, i.e. by neutrino oscillations and by the absorption of neutrinos on their way out of the core of the Sun.

\section{The Neutrino Spectrum}
\vspace{.2cm}
The annihilation of WIMPs generates neutrinos through a variety of channels. The annihilation products are heavy quarks, tau leptons, gauge bosons and/or Higgs bosons that decay into energetic neutrinos~\cite{Jungman:1995df}. In some models, WIMPs can also annihilate directly to neutrino-antineutrino pairs. Annihilations to light quarks or muons, however, do not contribute to the high energy neutrino spectrum because they are absorbed by the Sun before decaying.

Neglecting oscillations and the absorption of neutrinos in the Sun, the spectrum of neutrinos from WIMP annihilations to a final state, $X\bar{X}$, is given by:

\begin{equation}
\frac{dN_{\nu}}{dE_{\nu}} = \frac{1}{2} \int^{E_{\nu}/\gamma (1-\beta)}_{E_{\nu}/\gamma (1+\beta)}   \frac{1}{\gamma \beta}\frac{dE^{\prime}}{E^{\prime}} \bigg(\frac{dN_{\nu}}{dE_{\nu}}\bigg)^{\rm rest}_{XX},
\end{equation}
where $\gamma=m_{\chi}/m_X$, $\beta=\sqrt{1-\gamma^{-2}}$, and $(dN_{\nu}/dE_{\nu})^{\rm rest}_{XX}$ is the spectrum of neutrinos produced in the decay of an $X$ at rest.

As an example, consider the simple case of WIMPs annihilating to a pair of $Z$ bosons, $\chi \chi \rightarrow ZZ$. In this case, the most energetic neutrinos are produced by the direct decay of a $Z$ into a neutrino-antineutrino pair. In this case

\begin{equation}
\bigg(\frac{dN_{\nu}}{dE_{\nu}}\bigg)^{\rm rest}_{ZZ} = 2 \Gamma_{Z\rightarrow \nu \bar{\nu}} \, \delta(E_{\nu}-m_Z/2),
\end{equation}
which leads to
\begin{equation}
\bigg(\frac{dN_{\nu}}{dE_{\nu}}\bigg)_{ZZ} = \frac{2 \,\, \Gamma_{Z\rightarrow \nu \bar{\nu}}}{m_Z \, \gamma \, \beta}, \,\,\,\,\,\,\,\,\,\,\rm{for}\,\,\,\,\, m_{\chi}(1-\beta)/2 < E_{\nu} < m_{\chi}(1+\beta)/2,\,\,\,\,\,\,\,\,  
\label{zz}
\end{equation}
where $\Gamma_{Z\rightarrow \nu \bar{\nu}}\approx 0.067$ is the branching fraction to neutrino pairs (per flavor). In addition,  $Z$-bosons also produce neutrinos through their decays to $\tau^+ \tau^-$, $b\bar{b}$ and $c\bar{c}$.  An expression similar to Eq.~\ref{zz} can be written down for the case of WIMP annihilations to $W^+ W^-$, followed by $W^{\pm} \rightarrow l^{\pm} \nu$.

Tau leptons produce neutrinos through a variety of channels, including through the leptonic decays $\tau \rightarrow \mu \nu \nu$, $e \nu \nu$, and the semi-leptonic decays $\tau \rightarrow \pi \nu$, $K \nu$, $\pi \pi \nu$, and $\pi \pi \pi \nu$. Top quarks decay to a $W^{\pm}$ and a bottom quark nearly 100\% of the time, each can generate neutrinos in their subsequent decay. For bottom and charm quarks, only the semi-leptonic decays contribute to the neutrino spectrum (with the exception of the neutrinos resulting from taus and $c$-quarks produced in decays of $b$-quarks). Hadronization of $b$ and $c$ quark reduces the fraction of energy that is transferred to the resulting neutrinos and other decay products. To fully and accurately account for all such effects, programs such as PYTHIA are often used. 

Once produced in the Sun's core, neutrinos propagate through the solar medium before detection at Earth and may be absorbed and/or change flavor. In particular, absorption is the result of charged current interactions of electron and muon neutrinos in the Sun. The probability of absorption is given by $1-\exp(-E_{\nu}/E_{\rm abs})$, where $E_{\rm abs}$ is approximately 130 GeV for electron or muon neutrinos and 200 GeV for electron or muon antineutrinos. Absorption mostly affect the neutrinos in the case of relatively heavy WIMPs. For charged current intereactions of tau neutrinos, the tau leptons produced decay and thus regenerate a neutrino, albeit with a reduced energy. Neutral current interactions of all three neutrino flavors similarly reduce the neutrino's energy without depleting their number. 

Good sensitivity to neutrinos produced by WIMPS in the sun is achieved exploiting IceCube's degree accuracy with which secondary muon tracks, approximately aligned with the parent neutrino, can be pointed back to the Sun. The neutrinos oscillate before reaching the detector. Vacuum oscillations fully mix muon and tau neutrinos resulting into a muon neutrino spectrum that is the average of the muon and tau flavors prior to mixing. Electron neutrinos oscillate into muon flavor through matter effects in the Sun (the MSW effect). Electron antineutrinos can generally be neglected, as their oscillations to muon or tau flavors are highly suppressed~\cite{Lehnert:2007fv}. 

Program such as DarkSUSY~\cite{Gondolo:2005we}, which include effects such as hadronization, absorption, regeneration, and oscillations, are very useful in making detailed predictions for the neutrino spectrum resulting from WIMP annihilations in the Sun.

\section{Neutrino Telescopes}
\vspace{.2cm}
Muon neutrinos produce muons through charged current interactions with nuclei inside or near IceCube. The rate of neutrino-induced muons observed in a high-energy neutrino telescope is given by: 

\begin{eqnarray}
N_{\rm{events}} &\simeq& \int \int \frac{dN_{\nu_{\mu}}}{dE_{\nu_{\mu}}}\, \frac{d\sigma_{\nu}}{dy}(E_{\nu_{\mu}},y) \,[R_{\mu}(E_{\mu})+L]\, A_{\rm{eff}} \, dE_{\nu_{\mu}} \, dy  \nonumber \\
 &+& \int \int \frac{dN_{\bar{\nu}_{\mu}}}{dE_{\bar{\nu}_{\mu}}}\, \frac{d\sigma_{\bar{\nu}}}{dy}(E_{\bar{\nu}_{\mu}},y) \,[R_{\mu}(E_{\mu})+L]\, A_{\rm{eff}} \, dE_{\bar{\nu}_{\mu}} \, dy,
\end{eqnarray}
where $\sigma_{\nu}$ ($\sigma_{\bar{\nu}}$) is the neutrino-nucleon (antineutrino-nucleon) charged current interaction cross section, $(1-y)$ is the fraction of neutrino/antineutrino energy transferred to the muon, $A_{\rm{eff}}$ is the effective detection area of the detector, $R_{\mu}(E_{\mu})$ is the distance a muon of energy $(1-y)\,E_{\nu}$ travels before falling below the threshold $E^{\rm thr}_{\mu}$ of the experiment and $L$ is the depth of the detector volume linear size of a detector of volume $L^3$. The muon range in water/ice is well approximated by:

\begin{equation}
R_{\mu}(E_{\mu}) \approx 2.4 \, {\rm km} \,\times \, \ln\bigg[\frac{2.0+0.0042\,E_{\mu} ({\rm GeV})}{2.0+0.0042\,E^{\rm thr}_{\mu} ({\rm GeV})}\bigg].
\end{equation}
When completed, IceCube's instrumented volume will reach a L of one kilometer and an effective area of well over a full square kilometer for a threshold of approximately 100\,GeV. The Deep Core extension of Icecube will lower the threshold to 10\,GeV.

\begin{figure}
\begin{center}
\includegraphics[width=3.1in,angle=0]{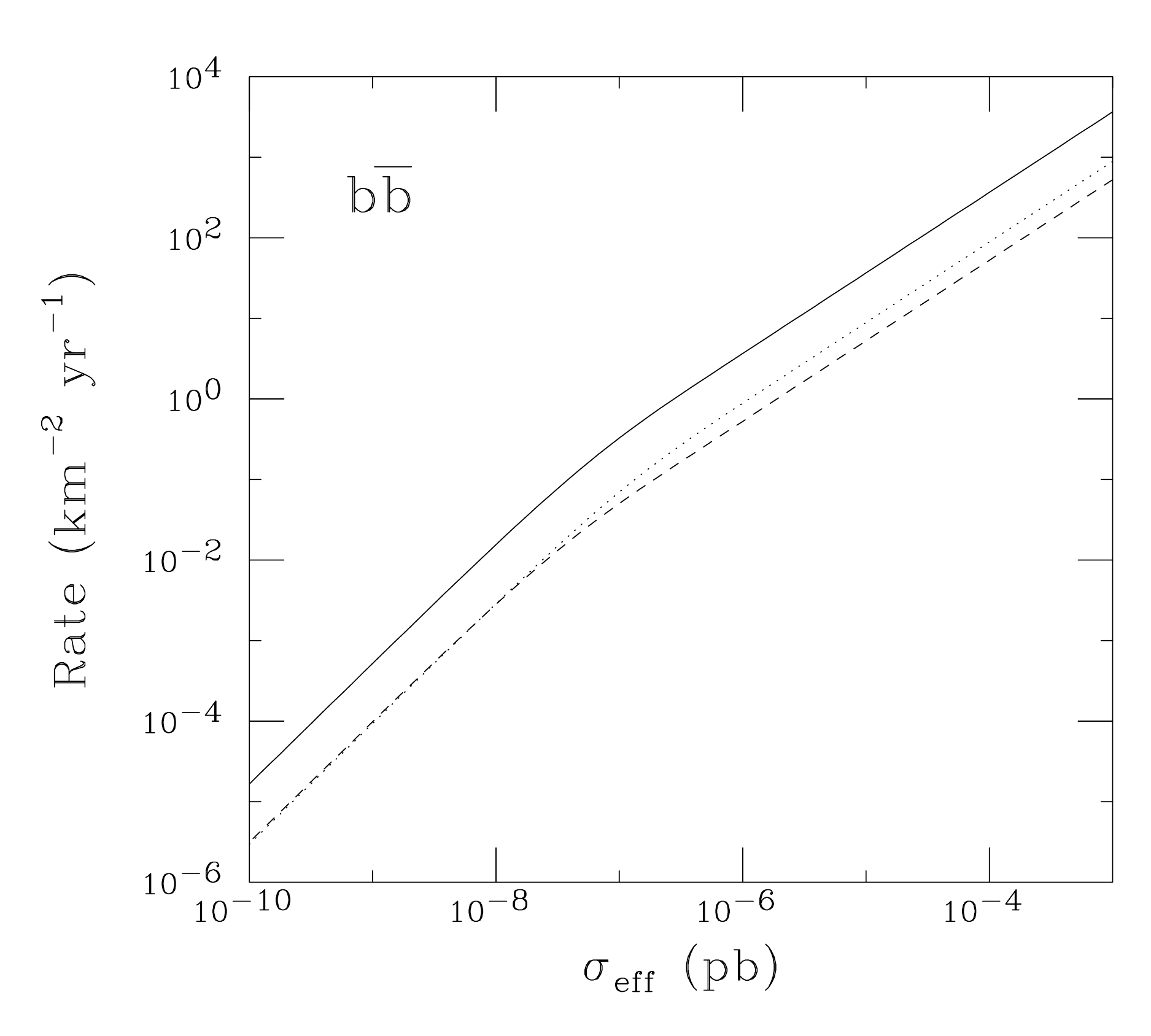}
\includegraphics[width=3.1in,angle=0]{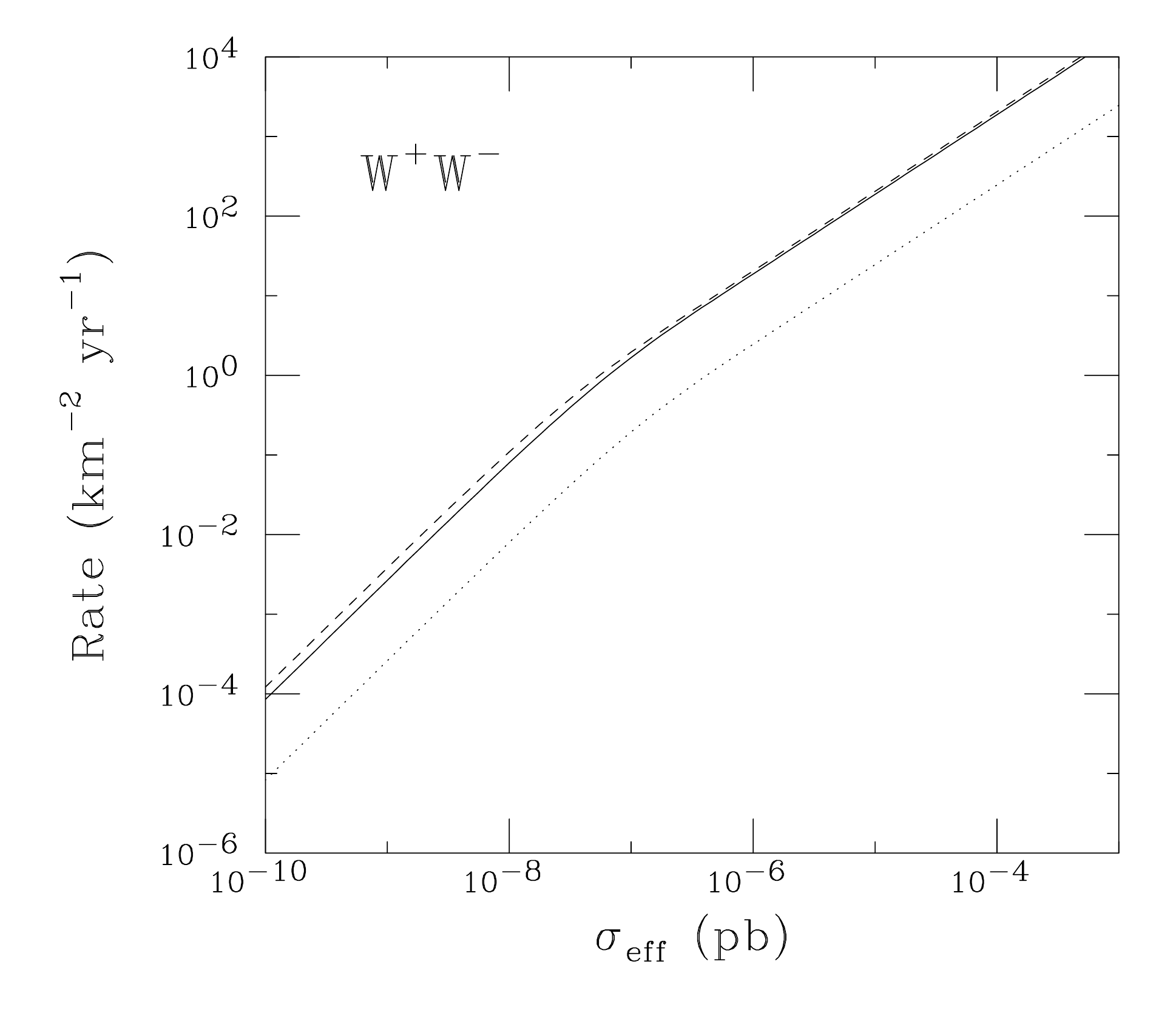}
\includegraphics[width=3.1in,angle=0]{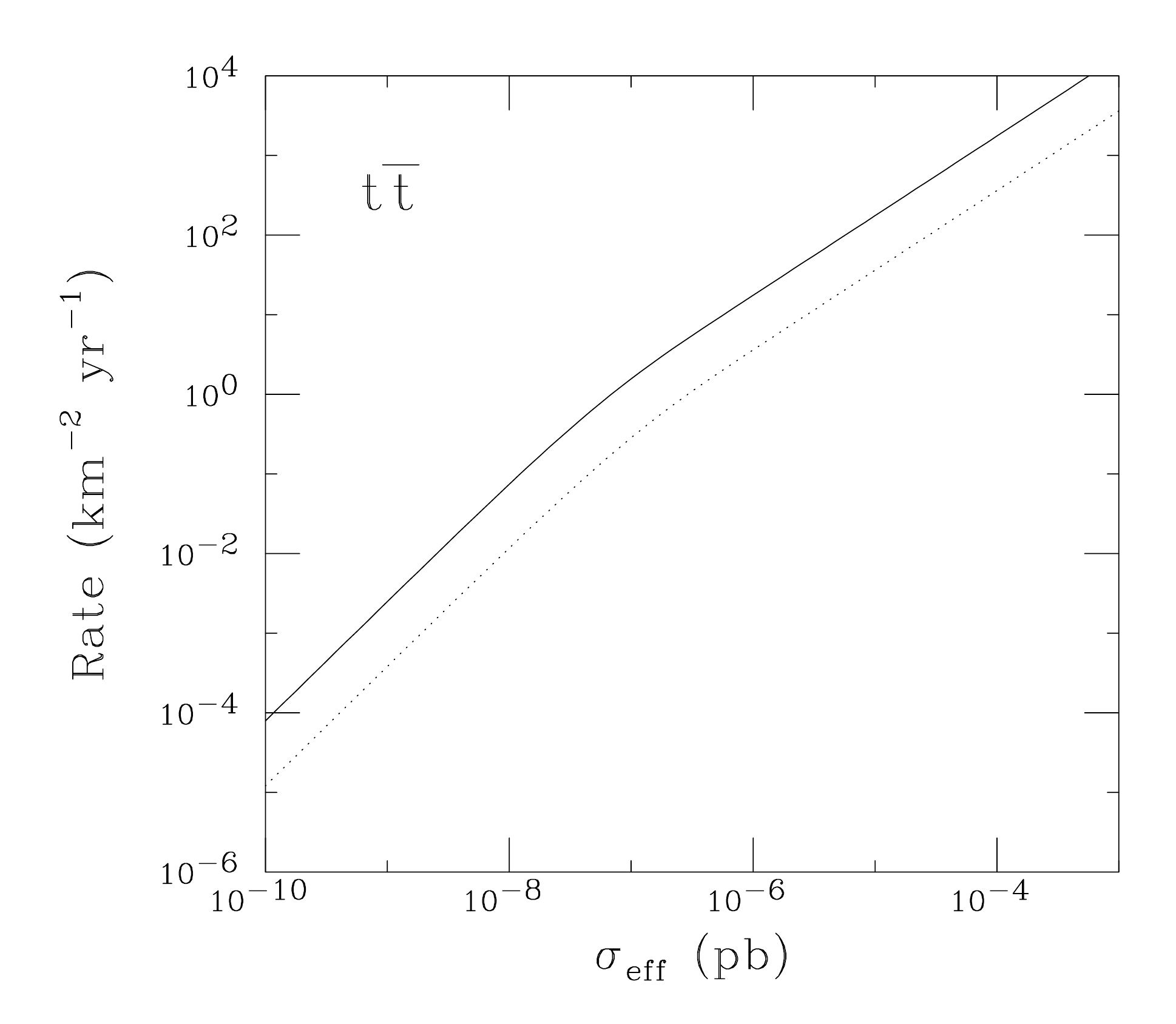}
\includegraphics[width=3.1in,angle=0]{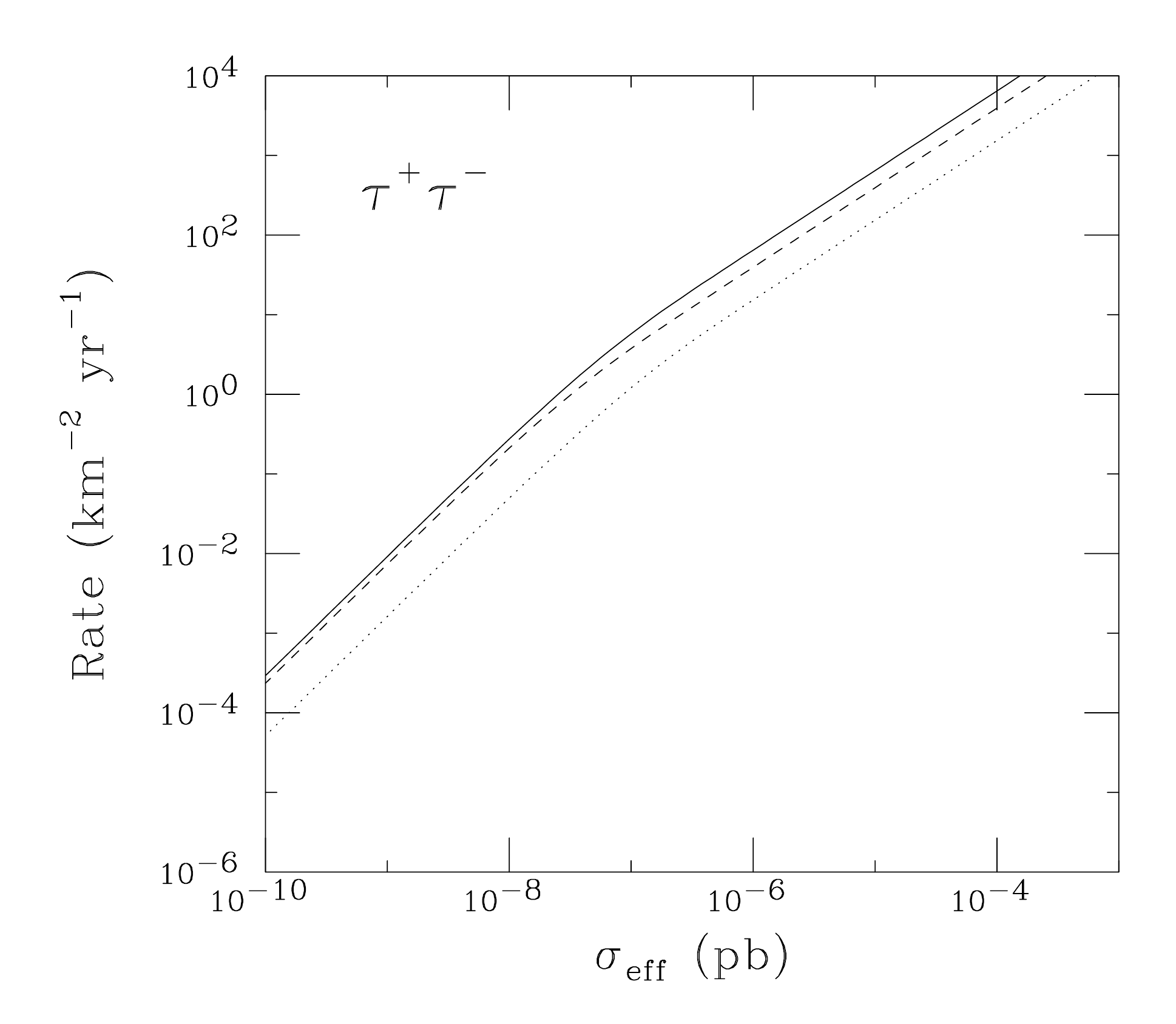}
\end{center}
\vspace{0cm}
\caption{The event rate in a kilometer-scale neutrino telescope such as IceCube as a function of the WIMP's effective elastic scattering cross section in the Sun for a variety of annihilation modes. The effective elastic scattering cross section is defined as $\sigma_{\rm{eff}} = \sigma_{\mathrm{\chi H, SD}} +\, \sigma_{\mathrm{\chi H, SI}} + 0.07 \, \sigma_{\mathrm{\chi He, SI}} + 0.0005 \, S(m_{\chi}/m_0) \, \sigma_{\mathrm{\chi 0, SI}}$ (see Eq.   ). The dashed, solid and dotted lines correspond to WIMPs of mass 100, 300 and 1000 GeV, respectively. A 50 GeV muon energy threshold and an annihilation cross section of $3 \times 10^{-26}$ cm$^{3}$ s$^{-1}$ have been adopted.}
\end{figure}

The flux and spectrum of the neutrinos from WIMP annihilation depend on the model parameters which determine the dominant annihilation channels. In Fig.\,4, we show the event rate in a kilometer-scale neutrino telescope such as IceCube as a function of the WIMP's elastic scattering cross section for four possible annihilation modes $b\bar{b}$, $t\bar{t}$, $\tau^+ \tau^-$ and $W^+W^-$. The effective elastic scattering cross section used here is defined as $\sigma_{\rm{eff}} = \sigma_{\mathrm{\chi H, SD}} +\, \sigma_{\mathrm{\chi H, SI}} + 0.07 \, \sigma_{\mathrm{\chi He, SI}} + 0.0005 \, S(m_{\chi}/m_0) \, \sigma_{\mathrm{\chi 0, SI}}$ (see Eq.~\ref{SIeff}).

The elastic scattering cross section of a WIMP is already constrained by the absence of a positive signal in direct detection experiments. Currently the strongest limits on the WIMP-nucleon spin-independent elastic scattering cross section have been obtained by the CDMS~\cite{Ahmed:2008eu} and XENON~\cite{Angle:2007uj} experiments. These results exclude spin-independent cross sections larger than approximately $5 \times 10^{-8}$ pb for a 25-100 GeV WIMP and $2 \times 10^{-7}$ pb ($m_{\chi}$/500 GeV) for a heavier WIMP.

With these results in mind, consider as an example a 300 GeV WIMP with an elastic scattering cross section with nucleons which is largely spin-independent. With a cross section near the CDMS bound, say $1 \times 10^{-7}$ pb, we obtain the corresponding signal in a neutrino telescope from Fig.\,4. Sadly, we find that this cross section yields less than 1 event per year for annihilations to $b\bar{b}$, about 2 events per year for annihilations to $W^+W^-$ or $t\bar{t}$ and about 8 per year for annihilations to $\tau^+ \tau^-$, none of which are sufficient to dominate the background of atmospheric neutrinos (see Fig.\,2). Clearly, WIMPs that scatter with nucleons mostly through spin-independent interactions are not likely to be detected with IceCube or other planned neutrino telescopes.

The state of affairs is diffferent for the case of spin-dependent scattering. The strongest bounds on the WIMP-proton spin-dependent cross section have been placed by the COUPP~\cite{Behnke:2008zza} and KIMS~\cite{Lee:2007qn} collaborations. These constraints are approximately 7 orders of magnitude less stringent than those for spin-independent cross sections. As a result, a WIMP with a largely spin-dependent scattering cross section with protons is capable of generating large event rates in high energy neutrino telescopes. For a 300 GeV neutralino with a cross section near the experimental limits indirect rates as high as $\sim$$10^3$ per year are expected; see  Fig.\,4.

\section{Neutralino Dark Matter}
\vspace{.2cm}
The elastic scattering and annihilation cross sections of the lightest neutralino depend on its couplings and on the mass spectrum of the Higgs bosons and superpartners. The couplings, in turn, depend on the neutralino's composition. In the Minimal Supersymmetric Standard Model (MSSM), the lightest neutralino is a mixture of the bino, neutral wino, and two neutral higgsinos: $\chi^0 = f_B \tilde{B} + f_W \tilde{W} + f_{H_1} \tilde{H_1} + f_{H_2} \tilde{H_2}$.

Spin-dependent, axial-vector, scattering of neutralinos with quarks and gluons is mediated by the t-channel exchange of a $Z$-boson, or the s-channel exchange of a squark. Spin-independent scattering occurs at the tree level through s-channel squark exchange and t-channel Higgs exchange, and at the one-loop level through diagrams involving loops of quarks and/or squarks.

The cross sections for these processes can vary dramatically with the parameters of the MSSM. A phenomenological description of the MSSM can be reduced to four mass parameters that determine the masses and couplings of the yet to be discovered supersymmetric particles and two more parameters that determine the properties of the higgs sector. These are $m_A$ the mass of the CP-odd higgs boson and $tan\beta$ the ratio of the vacuum expectation values of the other higgs bosons. By scanning the parameter space we create a set of models that will be used to evaluate the sensitivity of a kilometer-scale neutrino telescope. The scan varies all mass parameters up to 10 TeV, $m_A$ up to 1 TeV and the value of $tan\beta$ between 1 and 60. For generality, we do not assume any particular SUSY breaking scenario or unification scheme. Each model selected is consistent with all constraints for relevant collider experiments and with a thermal relic density not in excess of the value $\Omega_{\chi} h^2 =$0.129. We do not impose a lower limit on this quantity, keeping in mind the possibility of non-thermal processes which may contribute to the density of neutralino dark matter. We have performed the scan using the DarkSUSY program~\cite{Gondolo:2005we}. In Fig.\,5, we show the values of the spin-dependent and independent scattering cross sections as a function of the WIMP mass for all models generated by the scan.

\begin{figure}
\begin{center}

\includegraphics[width=3.1in,angle=0]{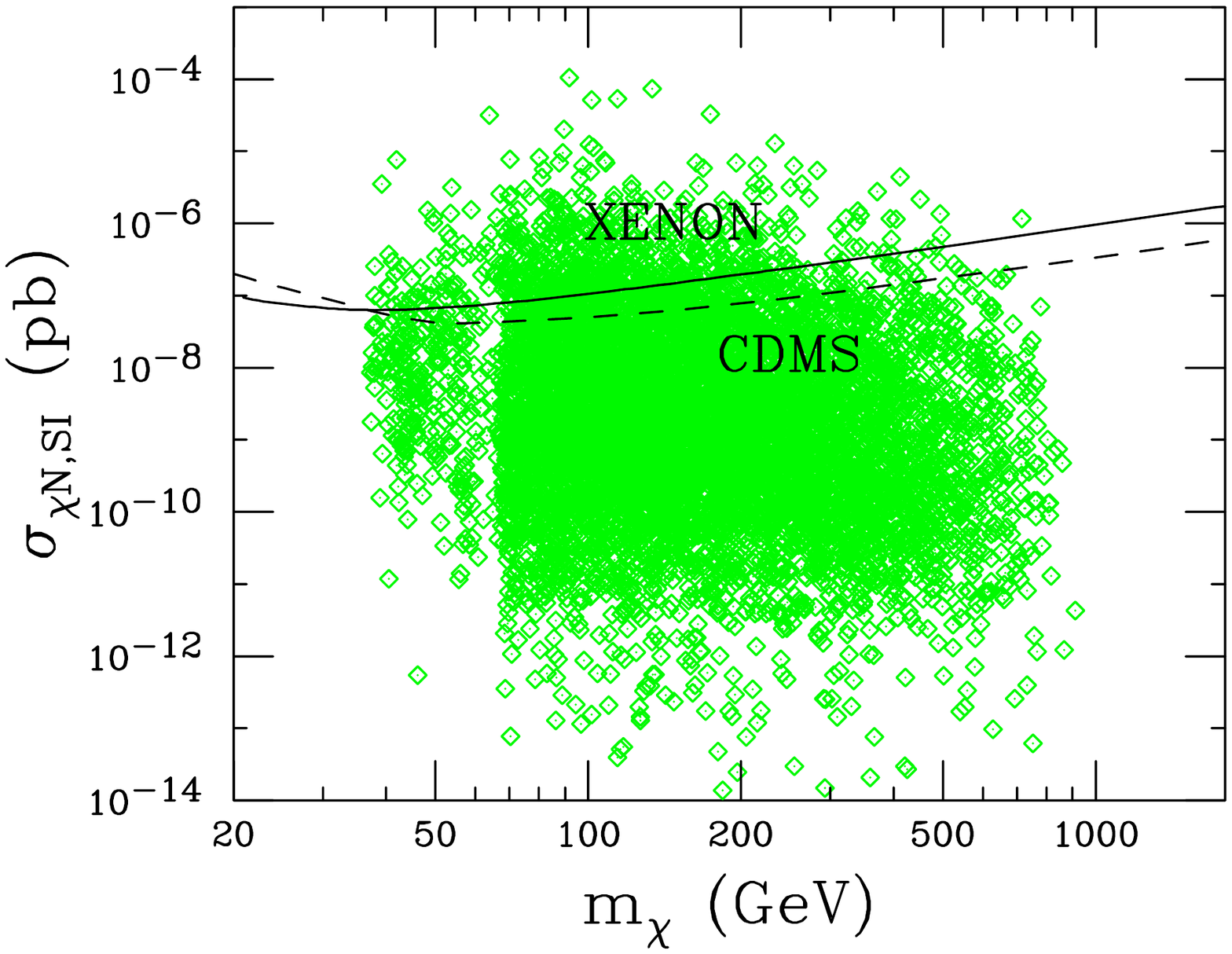}
\includegraphics[width=3.1in,angle=0]{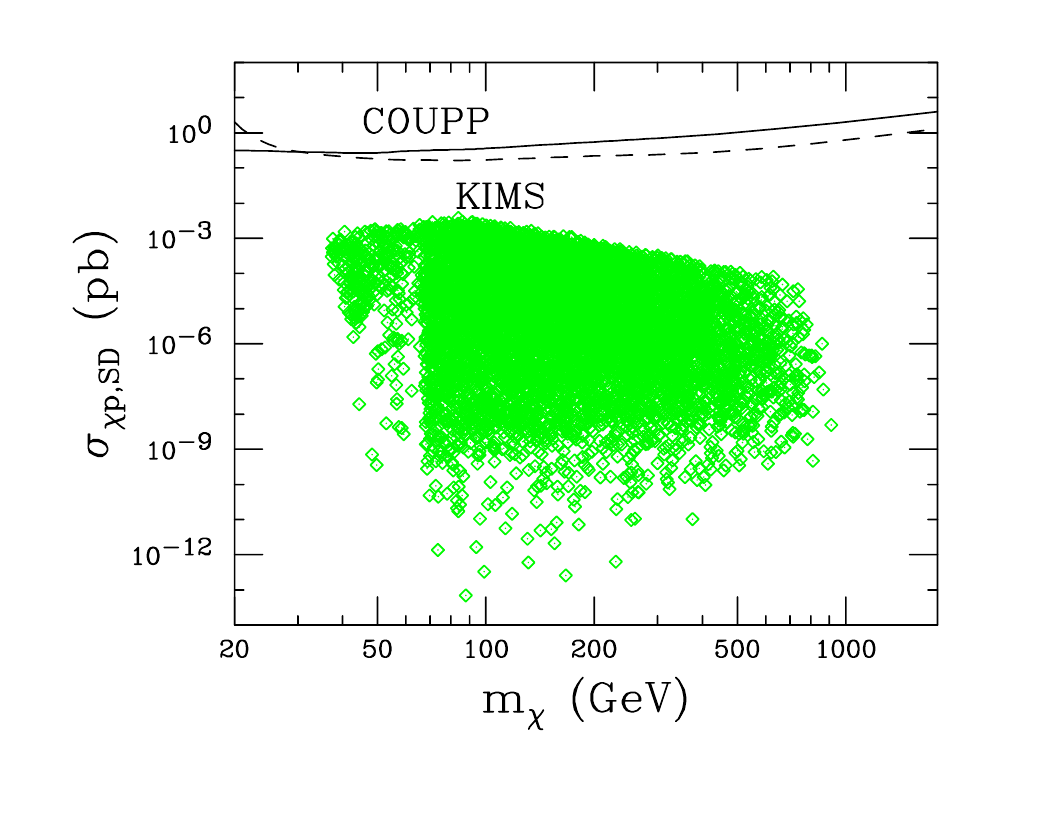}
\end{center} 

\caption{The lightest neutralino's spin-independent (left) and spin-dependent (right) scattering cross sections for a range of MSSM parameters. Also shown are the current limits from direct detection experiments.}

\end{figure}

The cross sections for these processes vary dramatically from model to model. It is clear from Fig.\,5, however, that for this class of models the spin-dependent cross section can be considerably larger than the spin-independent cross section, favoring indirect detection\cite{Ullio:2000bv}. In particular, very large spin-dependent cross sections ($\sigma_{\rm{SD}}\gtrsim10^{-3}$\,pb) are possible even in models with very small spin-independent scattering rates. Such a model would go easily undetected in all planned direct detection experiments, while still generating on the order of $\sim\,1000$ events per year at IceCube. In Fig.\,6, we quantify this point by plotting the event rate in a kilometer-scale neutrino telescope from neutralino annihilation in the Sun versus the neutralino's spin-dependent cross section with protons. In this figure, each point shown is beyond the reach of present and near future direct detection experiments. We thus conclude that neutralinos may be observable by IceCube while remaining beyond the reach of current and next generation direct detection experiments.

\begin{figure}
\begin{center}
\includegraphics[width=3.0in,angle=0]{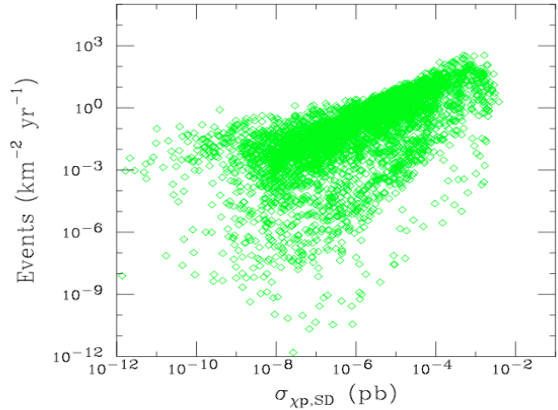}
\end{center}
\caption{The rate of events at a kilometer-scale neutrino telescope such as IceCube from neutralino dark matter annihilations in the Sun, as a function of the neutralino's spin-dependent elastic scattering cross section. Each point shown is beyond the reach of present and near future direct detection experiments.}
\end{figure}

A neutralino with a significant spin-dependent cross section generally has a sizable coupling to the $Z$ and, therefore, a large higgsino component. The spin-dependent scattering cross section through the exchange of a $Z$ is proportional to the square of the quantity $|f_{H_1}|^2 - |f_{H_2}|^2$. Even neutralinos with a higgsino fraction of a only a few percent are likely to be within the reach of IceCube~\cite{Bergstrom:1998xh,Halzen:2005ar}. This makes the focus point region of supersymmetric parameter space especially promising. In this region, the lightest neutralino is typically a strong mixture of bino and higgsino components, leading to the prediction of hundreds to thousands of events per year at IceCube. In Fig.\,7 we show that neutralinos with a higgsino component $|f_{H_1}|^2 - |f_{H_2}|^2$ of approximately one percent or greater are likely to be detectable by IceCube.

\begin{figure}
\begin{center}
\includegraphics[width=4.0in,angle=0]{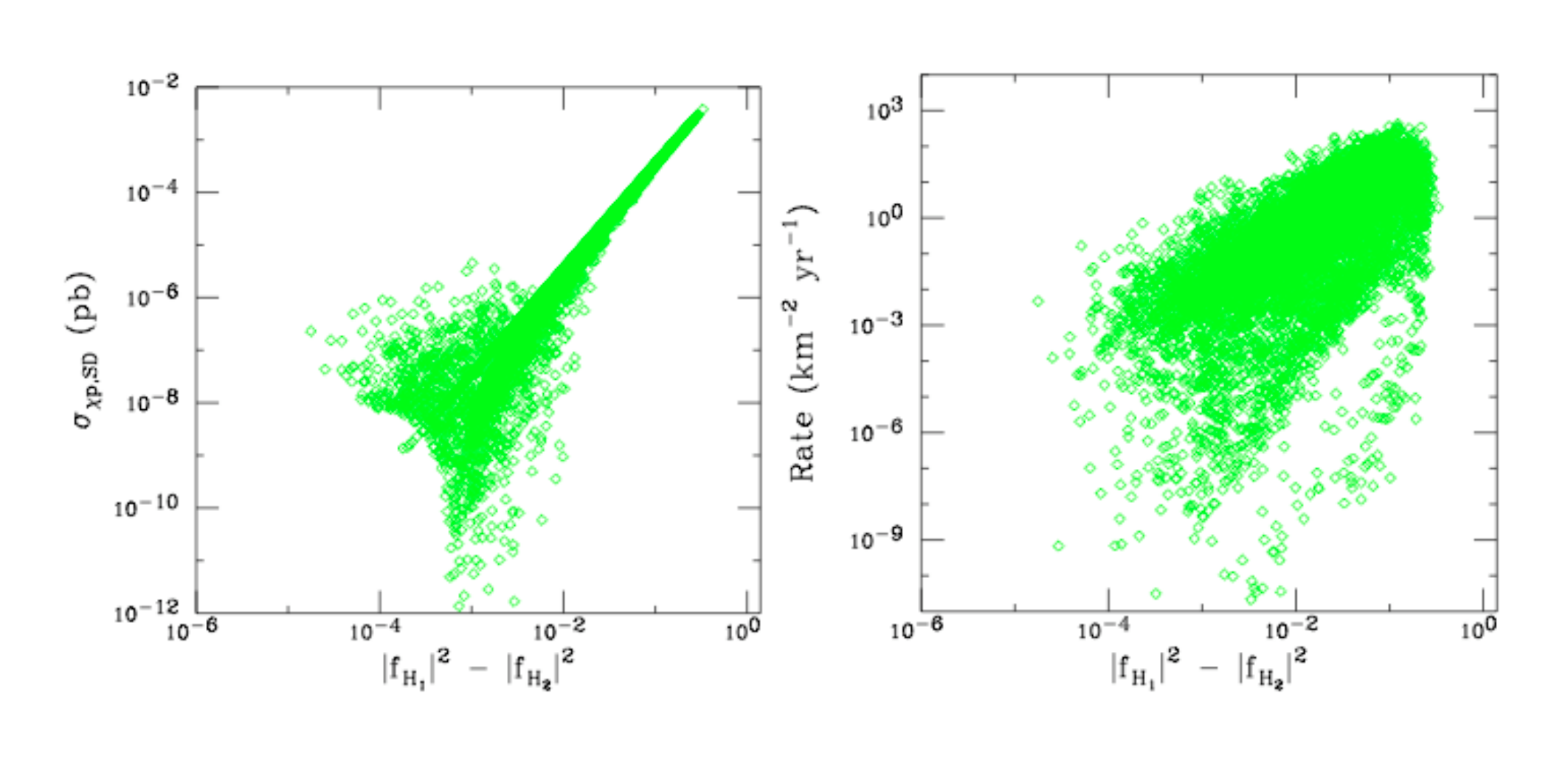}
\end{center}
\caption{In the left frame, the spin-dependent elastic scattering cross section of the lightest neutralino with protons is shown as a function of the quantity $|f_{H_1}|^2 - |f_{H_2}|^2$. In the right frame, the rate in a kilometer-scale neutrino telescope is shown, using a muon energy threshold of 50 GeV. Each point shown evades the current constraint of CDMS. See text for more details.}
\end{figure}

So far we have matched the prospects for the indirect detection of dark matter using neutrinos against a very general and relatively unconstrained set of supersymmetric models. Alternatively, several groups have made more restrictive scans of the MSSM parameter space. For instance, Allanach et al.~\cite{Allanach:2008iq} have identified islands in the $m_{\chi}$, $\sigma_{\chi p}$ space of models consistent with all empirical particle physics constraints. They are also constrained to yield the observed dark matter density of the Universe. Fig.\,8 shows the posterior probability for the spin-dependent cross section as a function of the neutralino mass for such a parameter scan. The definition of the posterior probability is not simple, we refer the reader to reference~\cite{Allanach:2008iq}. It represents the product of the likelihood and the prior, integrating over all parameters except for the ones that we are interested in. Regarding the normalization, the probability is defined relative to the maximum probability of any point in the parameter space. We conclude that the result is fairly robust because it turns out that the range of values for the spin-independent cross section is relatively insensitive to how the a priori likelihoods of points in the parameter space are defined. The side-by-side figures contrast scans using, alternatively, the parameters of the Higgs potential or $\tan\beta$ to define the parameter space.

It is well known that the MSSM generically generates values of the dark matter density which are in excess of the measured value. This is not surprising, as the ``WIMP miracle'' relating the weak force to $\Omega\sim1$ implies a generically weak cross section. Present searches probe WIMP cross sections smaller than the canonical weak-scale value by orders of magnitude. Allowed supersymmetric models are therefore limited to special regions in the parameter space where efficient annihilation can occur, for example through coannihilations or resonances. These special regions emerge in the scan shown in Fig.\,7 and include the focus point region (as previously mentioned), which is characterized by large spin-dependent cross sections in excess of $10^{-5}$\,pb.

\begin{figure}
\begin{center}
\includegraphics[width=6.0in,angle=0]{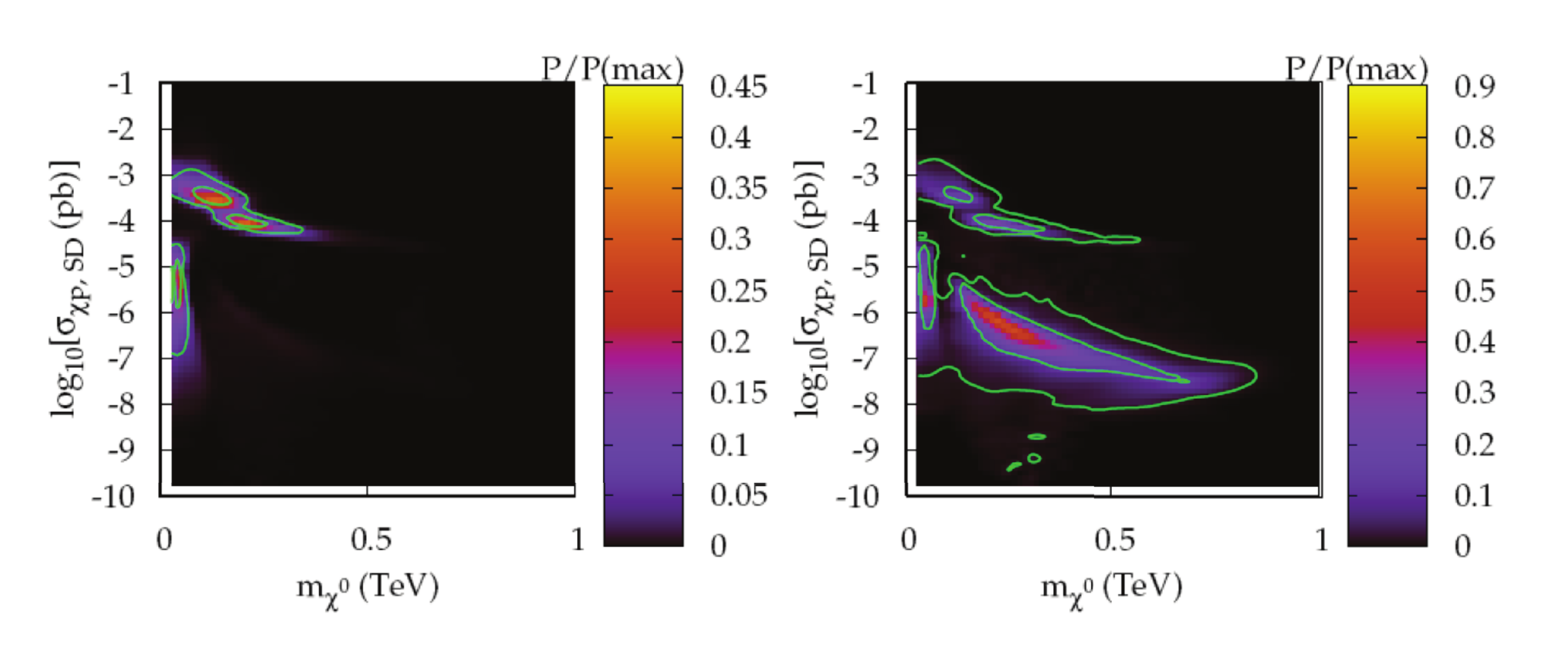}
\end{center}
\caption{Posterior probability distribution for the spin-dependent neutralino-proton elastic scattering cross section as a function of the neutralino mass using the higgs parameters (left) and $\tan\beta$ (right) to sample the model space. Contours delineate confidence regions of 68\% and 95\%.}
\end{figure}

For the models in the focus region, more than 60\% of the posterior probability distribution in the MSSM, we have converted the cross sections $\sigma_{\chi p}$ in Fig.\,7 into a flux of neutrinos from neutralino annihilation in the sun assuming a detector with a threshold of 50\,GeV. Because of the relatively large cross sections, the models typically yield more than a hundred high energy neutrino events per year from the direction of the Sun in a generic kilometer-size detector; see Fig.\,9. In fact, the present IceCube limit already excludes such models for the highest neutralino masses; see Fig.\,3. A second island of models clustered at very low masses corresponds to the so-called higgs pole region where efficient annihilation proceeds through the lightest CP-even higgs resonance into b and $\tau$ pairs. Exploring the set of models with low WIMP masses and cross sections below $10^{-5}$\,pb will be challenging and almost certainly require future megaton detectors.

\begin{figure}
\begin{center}
\includegraphics[width=3.1in,angle=0]{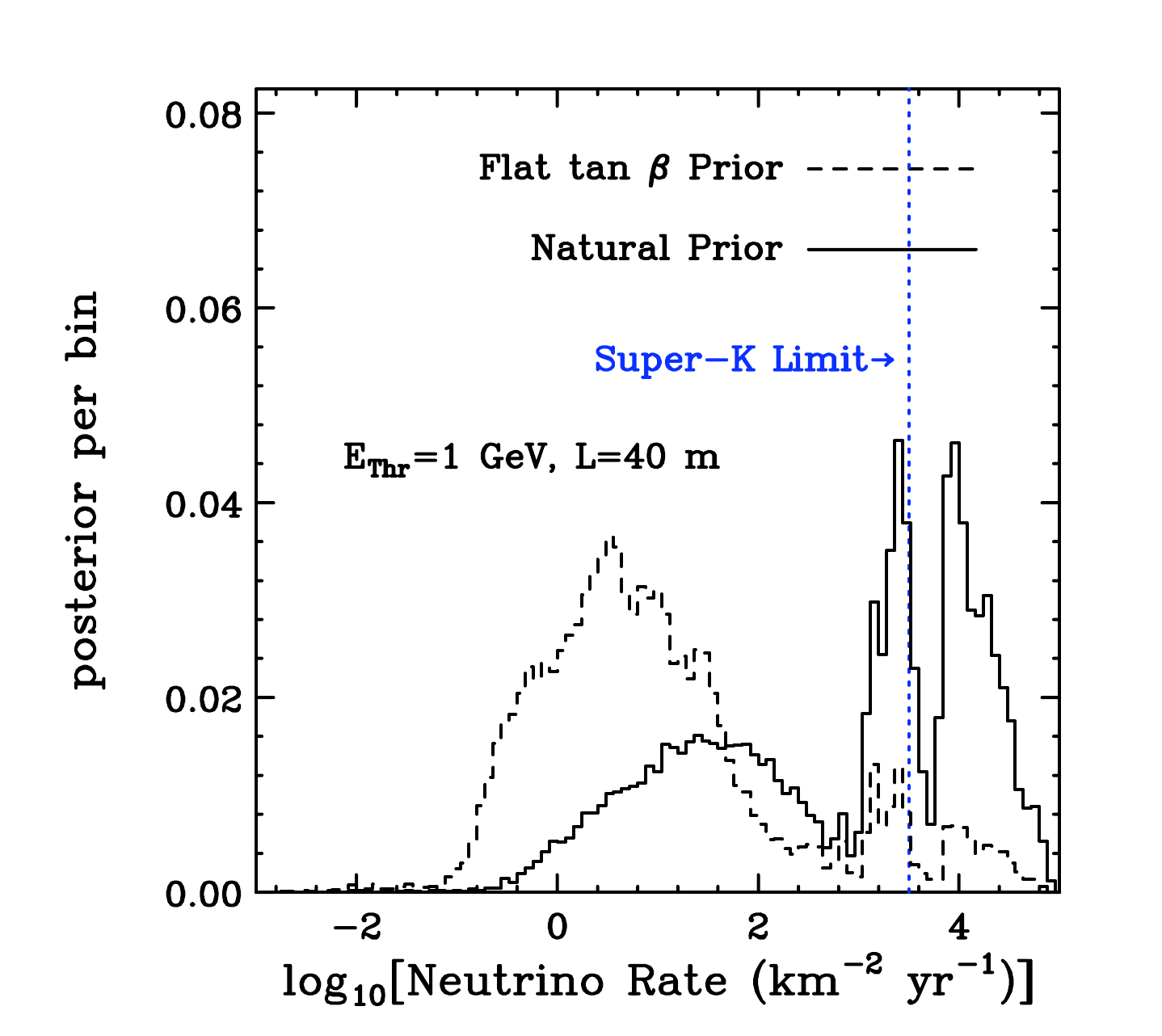}
\end{center}
\caption{Posterior probability distribution for the rate of neutrino events from neutralino annihilation in the Sun in a generic square kilometer detector with a threshold of 50\,GeV for the models shown in Fig.\,3.}
\end{figure}

This alternative look at the prospects for the indirect detection of dark matter using neutrinos confirms the excellent discovery potential derived from our previous less restrictive sample of minimal models.

\section{Alternative Dark Matter Models}
\vspace{.2cm}
A great variety of extensions of the Standard Model have been introduced which include new physics at or around the TeV-scale. A common feature of many of these scenarios is the inclusion of a symmetry which can stabilize the lightest new state, in much the same way that R-parity conservation stabilizes the lightest neutralino in supersymmetric models. Models with additional universal dimensions are of special interest in this context~\cite{Servant:2002aq,Cheng:2002ej}. In such models the fields of the Standard Model may propagate through the extended extra dimensional space, which is compactified on a small scale $R \sim \rm{TeV}^{-1}$. Each Standard Model particle is accompanied by Kaluza-Klein (KK) states which appear as heavy versions of their Standard Model counterparts with masses of $\sim 1/R$. The Lightest KK Particle (LKP) can be naturally stable and is likely to be the first KK excitation of the hypercharge gauge boson, $B^{(1)}$. This state can be produced with an abundance matching the observed dark matter density over a range of masses between 500\,GeV and a few TeV~\cite{Burnell:2005hm,Kong:2005hn}.

The spin-independent LKP-nucleon cross section turns out to be small and typically falls within the range of $10^{-9}$ to $10^{-12}$ pb~\cite{Servant:2002hb}, well beyond the sensitivity of current or upcoming direct detection experiments. On the other hand, the spin-dependent scattering cross section for the LKP with a proton is considerably larger~\cite{Servant:2002hb}:
\begin{eqnarray}
\sigma_{H, SD} &=& \frac{g_1^{4}\, m^2_p}{648 \pi m^4_{B^{(1)}} r^2_{q}} (4 \Delta_u^p + \Delta_d^p + \Delta_s^p)^2 \nonumber \\
&\approx& 4.4 \times 10^{-6} \,\rm{pb}\, \bigg(\frac{800 \,\rm{GeV}}{m_{B^{(1)}}}\bigg)^4 \, \bigg(\frac{0.1}{r_q}\bigg)^2,
\label{kkelastic}
\end{eqnarray}
where $r_q \equiv (m_{q^{(1)}}-m_{B^{(1)}})/m_{B^{(1)}}$ is fractional shift of the KK quark masses over the LKP mass, which is expected to be on the order of 10\%. The $\Delta$'s parameterize the fraction of spin carried by each variety of quark within the proton. In addition approximately 60\% of LKP annihilations generate a pair of charged leptons, 20\% to each flavor and about 4\% generate neutrino pairs. The neutrino and tau lepton final states each contribute substantially to the event rate in a neutrino telescope.

\begin{figure}
\begin{center}
\includegraphics[width=3.4in,angle=0]{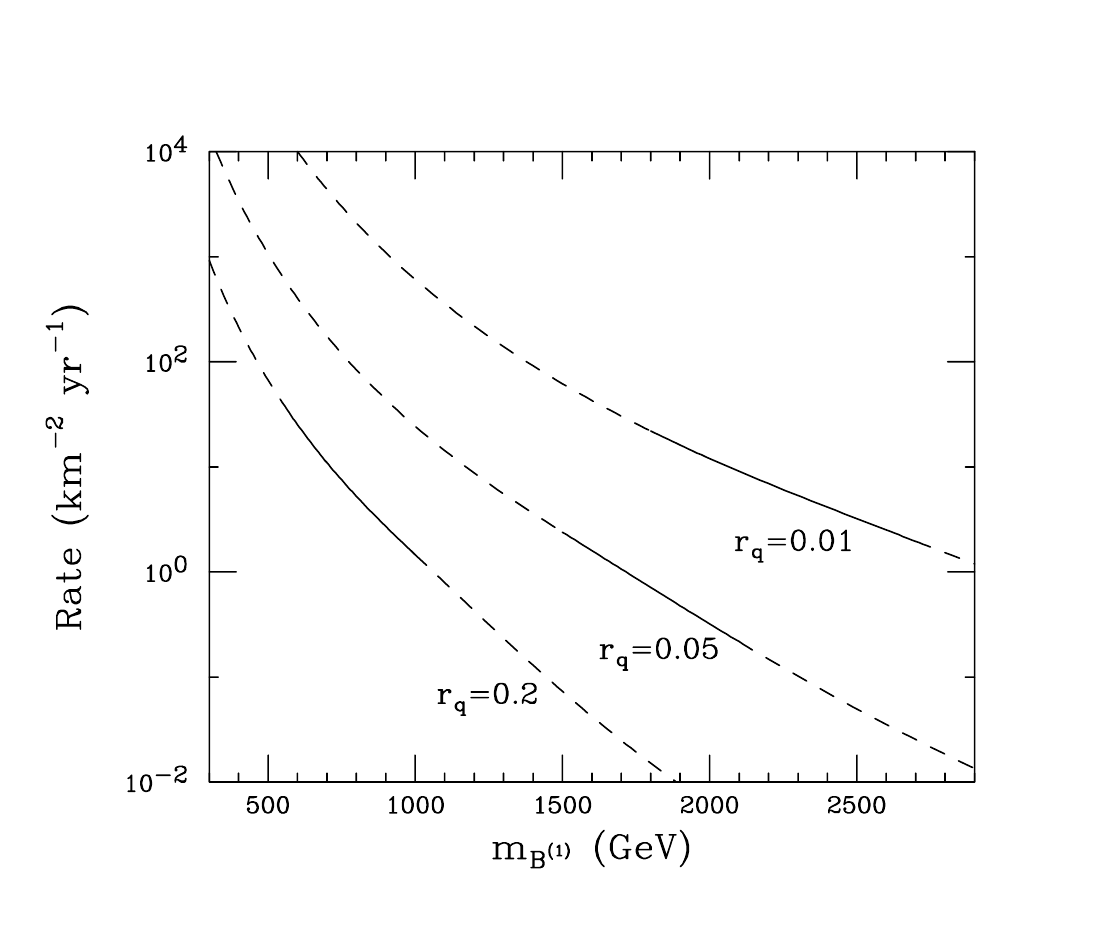}
\end{center}
\caption{The event rate in a kilometer-scale neutrino telescope as a function of the LKP mass~\cite{Hooper:2002gs}. The three lines correspond to fractional mass splittings of the KK quarks relative to the LKP of 20\%, 5\% and 1\%. The solid sections of these lines reflect the approximate range in which it is possible to generate the observed thermal relic abundance. A 50 GeV muon energy threshold has been used.}
\end{figure}

The event rates in a kilometer scale neutrino telescope for KK dark matter annihilating in the Sun~\cite{Hooper:2002gs} are shown in Fig.\,10. There are competing effects which contribute to these favorable results. In particular, a small mass splitting between the LKP and KK quarks yields a large spin-dependent elastic scattering cross section; see Eq.~\ref{kkelastic}. On the other hand, KK quarks which are not much heavier than the LKP contribute to the freeze-out process and increase the range of LKP masses in which the thermal abundance matches the observed dark matter density. For this range shown as the solid line segments in Fig.\,10, between 0.5 and 50 events per year are expected in a kilometer scale neutrino telescope such as IceCube. Alternatively, if non-thermal mechanisms, such as decays of KK gravitons, contribute to generating the LKP relic abundance, much larger rates are possible.

\end{document}